\documentstyle[preprint,aps,floats,epsfig]{revtex}

\newcommand{\be}{\begin{equation}}
\newcommand{\ee}{\end{equation}}
\newcommand{\bea}{\begin{eqnarray}}
\newcommand{\ea}{\end{eqnarray}}
\newcommand{\bml}{\begin{mathletters}}
\newcommand{\eml}{\end{mathletters}}

\begin{document}

\tighten

\preprint{DCPT-02/73}
\draft




\title{Quasi exactly solvable operators and Lie superalgebras}
\renewcommand{\thefootnote}{\fnsymbol{footnote}}
\author{ Yves Brihaye\footnote{Yves.Brihaye@umh.ac.be}}
\address{Facult\'e des Sciences, Universit\'e de Mons-Hainaut,
 B-7000 Mons, Belgium}
\author{Betti Hartmann\footnote{Betti.Hartmann@durham.ac.uk}}
\address{Department of Mathematical Sciences, University
of Durham, Durham DH1 3LE, U.K.}
\date{\today}
\setlength{\footnotesep}{0.5\footnotesep}

\maketitle
\begin{abstract}
Linear operators preserving the direct sum of polynomial 
rings ${\cal P}(m) \oplus {\cal P}(n)$ are constructed. 
In the case $\vert m-n \vert = 1$ they correspond to
atypical representations of the superalgebra osp(2,2).
 For $\vert m-n \vert = 2$
the generic, finite dimensional representations of the 
superalgebra q(2) are recovered. An example of a Hamiltonian
possessing such a hidden algebra is analyzed.
\end{abstract}


\renewcommand{\thefootnote}{\arabic{footnote}}

\section{Introduction}
Quasi exactly solvable (QES) operators \cite{tur} refer to linear 
differential operators which preserve a finite dimensional
vector space of smooth functions. They can be used in the framework
of quantum mechanics to construct Hamiltonians which possess a finite
number of algebraic eigenfunctions. The classification
of QES operators constitutes an interesting mathematical problem
which generalizes the Bochner problem \cite{tur2}.

Having chosen the vector space ${\cal V}$
of functions to be left invariant,
the first trial is to construct a set of basic operators preserving
${\cal V}$ from which all others can be generated in the 
sense of an enveloping algebra. The fact of having a set of normal
ordering rules for the basic operators is also crucial at this stage.

However, one of the striking aspects of QES operators is their close
relation to the theory of representations of Lie algebras and
superalgebras. In particular the realisations of Lie algebras
in terms of differential operators play a crucial role \cite{bgk}.

Of course, the fact that QES operators are particular realisations
of Lie algebras is not a necessary requirement, as shown in \cite{bk}, but it 
constitutes an advantage because, if so, representation
theory helps constructing families of invariant vector spaces.
In this paper, we put emphasis on the realisations of the
algebra q(2) in terms of QES operators. We discuss some generalities in
Section II and give our construction in Section III. 
An example of a QES Schr\"odinger equation related to these operators
is analyzed in Section IV. 
We end with concluding remarks and
an outlook in Section V.

\section{Generalities}
Let us denote by ${\cal P}(n)$ the ring of polynomials of degree less or equal to $n$
in a real variable $x$. The main result concerning the 
construction of QES operators \cite{tur} is the discovery that
the linear operators which preserve ${\cal P}(n)$ constitute
the enveloping algebra of the 
Lie algebra sl(2) in the representation
\begin{equation}
j^+_n = x(D-n)\ ,\ j^0_n = D-{n\over 2}\ ,\ j^- = {d\over{dx}} \ \ 
{\rm with} \ \  D\equiv x{d\over{dx}}.
\end{equation}
 
Extending this result to the case of
linear operators which preserve the vector
space ${\cal P}(n-\Delta)\oplus {\cal P}(n)$
it was shown in \cite{bk} that these operators can be assembled from
$2 \Delta + 6$  
basic operators  which can be chosen according to 
\begin{equation}
\label{T}
T^+ = 
\left(\begin{array}{cc}
j^+_{(n-\Delta)} &0\\
0 &j^+_{n}
\end{array}\right)
\quad , \quad 
T^0 =
\left(\begin{array}{cc}
j^0_{(n-\Delta)} &0\\
0 &j^0_{n}
\end{array}\right)
\quad , \quad 
T^- =
\left(\begin{array}{cc}
j^- &0\\
0 &j^-
\end{array}\right)
\end{equation}
\begin{equation}
\label{J}
J = {1\over 2} 
\left(\begin{array}{cc}
n+\Delta &0\\
0 &n  \ .
\end{array}\right)
\end{equation}
The three operators $T^{\pm,0}$ obey the $sl(2)$ algebra :
\begin{equation}
[T^+,T^-] = -2T^0\ ,\ [T^{\pm},T^0] = \mp T^{\pm}
\end{equation}
and $J$ commutes with all $T$'s.  
In the following these operators play the role of
bosonic operators. They have to be 
completed by $2 (\Delta + 1)$ off-diagonal ones~:
\begin{equation}
\label{Q}
Q_{\alpha} = x^{\alpha-1}\sigma_-\qquad 
\alpha = 1 , \cdots , \Delta + 1
\end{equation}
\begin{equation}
\label{Qb}
\bar Q_{\alpha} = \bar q_{\alpha, (n)}\sigma_+
\quad , \quad \alpha = 1, \cdots , \Delta + 1
\end{equation}
with the definition
\begin{equation}
\bar q_{\alpha,(n)} = 
\Bigl(
\prod_{j=0}^{\Delta - \alpha}(D - (n+1-\Delta)-j) 
\Bigr)
(\frac{d}{dx})^{\alpha-1}
\end{equation}
and  $\sigma_{\pm}= (\sigma_1 \pm i \ \sigma_2)/2$.

\subsection{Commutation relations}
After an algebra, the commutation relations between the
diagonal operators and the off-diagonal ones can be obtained \cite{bk}~:
 \begin{eqnarray}
\lbrack T^+, Q_{\alpha}\rbrack &=& -(1-\alpha+\Delta) Q_{\alpha+1}\\
\lbrack T^0, Q_{\alpha}\rbrack &=& -(1-\alpha+\frac{\Delta}{2})Q_{\alpha}\\
\lbrack T^-, Q_{\alpha}\rbrack &=& -(1-\alpha) Q_{\alpha-1} 
\end{eqnarray}
\begin{eqnarray}
\lbrack T^+, \overline Q_{\alpha}\rbrack &=& (1-\alpha) 
\overline Q_{1-\alpha}\\
\lbrack T^0, \overline Q_{\alpha}\rbrack &=& (1-\alpha+\frac{\Delta}{2}) \overline Q_{\alpha}\\ 
\lbrack T^-, \overline Q_{\alpha}\rbrack &=& (1-\alpha+ \Delta)  
\overline Q_{\alpha+1}
\end{eqnarray}
Using the  representations of sl(2), these formulae reveal that the set 
of operators $Q$ (and independently $\overline Q$) transforms
according to the representation of spin $s\equiv \frac{\Delta}{2}$
under the adjoint action of  $T$'s (they are also called tensorial operators).
 In fact both $\overline Q_{\alpha}$ and
$P_{\alpha} \equiv Q_{\Delta + 2 - \alpha}$ behave exactly the same
under the $T$'s. 
In terms of Young diagrams, the representation
generated by the $\overline Q$'s (or the $P$'s) is characterized
by the Young diagram with one line of $\Delta$ boxes.

 On the other hand, $J$ behaves as a grading operator~:
\begin{equation}
\lbrack J , Q_{\alpha} \rbrack = - \frac{\Delta}{2} Q_{\alpha} \ \ , \ \
\lbrack J , \overline Q_{\alpha} \rbrack 
=  \frac{\Delta}{2} \overline Q_{\alpha}
\end{equation}

One can then convince oneself that the most natural set 
of ordering rules
is obtained when {\it anticommutators} between the $Q$'s and the $\overline Q$'s
are chosen, which makes the interpretation of the $Q$s and $\overline Q$s
as fermionic operators natural.
The unpleasant feature about the algebraic structure generated by 
$T, J, Q, \overline Q$ is that the anticommutators 
$\{Q_{\alpha} , \overline Q_{\beta} \}$
are generically  polynomials of degree $\Delta$ in the
bosonic operators. In the case $\Delta = 1$ the operators
constitute an atypical representation of the superalgebra
osp(2,2) \cite{bgk}.
For $\Delta > 1$ the operators $T, Q, \overline Q$ do not seem to be related to
Lie superalgebras.
However, we will show in the following section
that it is possible to find a relation  for $\Delta = 2$.

\section{The case $\Delta = 2$}
We now consider the case $\Delta = 2$. This case has the peculiarity that, under the $T$'s,
the operators $Q$ and the $P$ transform according to the adjoint representation, 
in other words they transform as triplets of sl(2). Let us characterize a triplet
$V_1,V_2,V_3$ by
\begin{eqnarray}
\lbrack T^+,V_{\alpha}\rbrack &=& (1-\alpha) V_{\alpha-1} \nonumber \\
\lbrack T^0, V_{\alpha}\rbrack &=& (2-\alpha)V_{\alpha} \\ 
\lbrack T^-,V_{\alpha}\rbrack &=& (3-\alpha) V_{\alpha+1} \nonumber
\end{eqnarray}
with $\alpha = 1,2,3$.
 The following sets of operators 
\begin{eqnarray}
\bar Q_{\alpha} &\equiv& (\overline Q_1, \overline Q_2, \overline Q_3)\\
P_{\alpha} &\equiv& (Q_3, Q_2, Q_1)\\
T_{\alpha} &\equiv& (T^+, T^0, T^-)\equiv (T_1, T_2, T_3)
\end{eqnarray}
therefore transform  as triplets under the $T$'s. Obviously
any linear combination of these triplets (with coefficients
being operators commuting with the $T$'s)  also constitutes a triplet.
 
The case $\Delta=2$ in this sense in special since for 
$\Delta > 2$ no match
between the fermionic and the bosonic operators seems to exist.
This is also true for QES operators depending on many variables
\cite{bn}. In this case the diagonal operators
obey the commutation
relations of sl(V+1) with V being the number of 
independent variables. The
counterpart of the $Q$'s then corresponds to the 
completely symmetric representation of sl(V+1) with
a Young diagram containing one line with $\Delta$ boxes. In contrast, the adjoint
representation of sl(V+1) is characterized by a diagram with
$V-1$ lines, the first line with two boxes, the others with one box.
The case $V=1$, $\Delta = 2$ is therefore very peculiar.

To proceed further we remind that in the case studied here
the anticommutators of $Q$ with $\bar Q$
lead to quadratic polynomials in the $T's$ and $J$, 
the details of which are given in \cite{bk}.
In order to obtain a more conventional algebra, we take advantage of 
the coexistence of three independent triplets of operators and 
try to simplify the algebra of operators preserving ${\cal P}(n-2) \oplus {\cal P}(n)$. 
For this purpose,
we define a new triplet of operators, $F_{\alpha}$~: 
\begin{equation}
\label{combi}
F_{\alpha}\equiv \bar Q_{\alpha} + cP_{\alpha}+DT_{\alpha}\quad , \quad \alpha=1,2,3
\end{equation}
where $c$ is a constant and $D$ is a constant diagonal matrix. 
If we choose $D^2={\bf{1}}_2$ and $c=-1$
we obtain
\begin{equation}
\label{linear}
\lbrace F_{\alpha},F_{\beta}\rbrace = n^2 g_{\alpha\beta} \ ,
\end{equation}
where $g_{\alpha \beta}$ is the Cartan metric of sl(2).
In the representation
used, the non-zero elements of this metric are given by
\begin{eqnarray}
g_{\alpha\beta} &=& 1\  \ \ \ {\rm{if}}\ \ \  \alpha , \beta = 1 , 3\ {\rm{or}}\ 3 , 1\\
&=& -{1\over 2} \ {\rm{if}}\  \ \  \alpha=\beta=2
\end{eqnarray}
From (\ref{linear}) it is apparent that
the anticommutators $\lbrace F_{\alpha}, F_{\beta}\rbrace$  are now linear combinations of the bosonic operators.

To establish this result, the identities
\begin{eqnarray}
j_{\beta, (n)} p_{\alpha}-p_{\alpha}j_{\beta, (n-2)} &=& (\beta-\alpha) p_{\alpha+\beta-2}\\
j_{\beta, (n-2)}\bar q_{\alpha, (n)}-\bar q_{\alpha, (n)} j_{\beta, (n)} 
&=& (\beta-\alpha) \bar q_{\alpha+\beta-2, (n)}
\end{eqnarray}
have to be used, and  we  introduced the notations
 $p_{\alpha} \equiv x^{3-\alpha}$ and 
$j_{1, (n)} \equiv j^+_n, j_{2, (n)} \equiv j^0_n, j_{3, (n)} \equiv j^-$.

As an alternative to the operators (2)-(6) above,
 we propose the  set given by $T_1 , T_2 , T_3$ and completed by
\begin{equation}
\hat F_{\alpha} \equiv {1\over{\sqrt{n}}} F_{\alpha} \ \ ,\alpha=1,2,3
\ \ , \ \ h_0 = n {\bf{1}}_2 \ \ , \ \   h_1 = \sqrt n \sigma_3  \ .
\end{equation}
Now, $T^{\pm,0}, h_0$ are the bosonic operators and $\hat F_{\alpha}, h_1$
the fermionic operators. They fulfill the (anti)-commutation relations
of the superalgebra q(2) \cite{dv}. In particular, we find 
\begin{eqnarray}
\lbrace \hat F_{\alpha}, \hat F_{\beta}\rbrace &=& g_{\alpha\beta}h_0\\
\lbrace \hat F_{\alpha}, h_1 \rbrace &=& 2T_{\alpha}\\
\lbrace h_1,h_1 \rbrace &=& 2 h_0
\end{eqnarray}
This latter set of operators therefore constitutes a series of
realisations of the superalgebra q(2) by QES operators. This
series is labelled by an integer $n$ and the 
$2n$ dimensional vector space preserved
is ${\cal P}(n-2) \oplus {\cal P}(n)$. Notice that this result
was presented in \cite{dv}, but the calculation was done by ``brute
force". Our derivation uses the representation
structure of the operators $Q$ and $\overline Q$ and moreover demonstrates
the importance of the case $\Delta = 2$.

\section{Example of a QES Hamiltonian for $\Delta=2$}
\par In this section, we discuss examples of  QES Schr\"odinger operators preserving the vector space 
${\cal P}(n)\oplus {\cal P}(n-2)$ and construct the eigenvalues for $n=2$ and $n=3$. The Hamiltonian is given by:
\begin{equation}
\label{ham}
H = -{d^2\over{dy^2}}{\bf{1}}_2 +y^6{\bf{1}}_2 + (1-4n)y^2{\bf{1}}_2-4y^2\sigma_3-4nk_0\sigma_1
\end{equation}
where $\sigma_1,\sigma_3$ are the Pauli matrices, $k_0$ is 
an arbitrary constant and $n$ an integer.
To our knowledge, this Hamiltonian is the only possible QES matrix Hamiltonian with polynomial potential 
\cite{zh,bh} ( which can however
be generalised through the inclusion of
a $y^4$ term). We introduce the following  change of basis and variable:
\begin{equation}
\hat H = U^{-1} HU\  \ \ {\rm with} \ \  \ U=e^{-\frac{x^2}{4}} \left(\begin{array}{cc}
1 &0\\
k_0{d\over{dx}} &1
\end{array}\right)\quad , \quad x=y^2
 \ .
\end{equation}
Then, the Hamiltonian reads:
\begin{eqnarray}
\hat H &=& -(4 x {d^2\over{dx^2}} + 2{d\over{dx}}){\bf{1}}_2-4nk^2_0{d\over{dx}} \sigma_3
+ 4
\left(\begin{array}{ll}
x^2{d\over {dx}}-nx & 0\\
0 & x^2{d\over{dx}}-(n-2)x
\end{array}\right)\nonumber\\
&+&4k_0
\left(\begin{array}{ll}
0 & -n\\
(1+k^2_0n){d^2\over{dx^2}} & 0
\end{array}\right)
\end{eqnarray}
which obviously preserves ${\cal P}(n)\oplus {\cal P}(n-2)$.
\par As a consequence, $2n$ algebraic eigenvectors of $\hat{H}$ can be 
constructed. In the limit $k_0=0$ two decoupled sextic QES Hamiltonians are recovered.
In the following, we discuss the algebraic spectrum of $\hat{H}$ for $n=2$ and $n=3$. 

For $n=2$ the four eigenvalues are given by 
\begin{equation}
E^2 = 32 + c^2\pm 4\sqrt{64+2c^2}\quad , \quad c \equiv -4n k_0
\end{equation}
and the corresponding eigenvectors are 
\begin{equation}
\label{eigenv}
\phi \propto
\left(\begin{array}{c}
y^4 - {E\over 4} y^2+{E^2-c^2-48\over{32}}\\
-{c\over 4} y^2 + {E(E^2-c^2-64)\over{32c}}
\end{array}\right)
e^{-{y^4\over 4}}
\end{equation}

The degeneracy of the energy levels for $c=0$ with $E=-8,0,0,8$ 
is lifted for $c>0$. 
In the limit $c \rightarrow 0$, two
of the eigenvectors (\ref{eigenv}) converge to linear combinations
of the two eigenvectors of the decoupled system with $E=0$.
The ground state energy corresponds to $E_{(0)}=-\sqrt{32+c^2+4\sqrt{64+2c^2}}$
and both  components of the corresponding eigenvector 
$\phi$ have no nodes. 
Denoting the four algebraic energy levels by $E_{(a)}$, $a=0, 1, 2, 3$, 
in increasing order and by $k_{1(a)}$, $k_{2(a)}$ the number of nodes
of the two components of the corresponding eigenvector $\phi_{(a)}$, we obtain:
\begin{center}
Table 1 \\
\medskip
\begin{tabular}{|l|l|l|l|l|}
\hline 
$ $ & $a=0$ & $a=1$ & $a=2$ & $a=3$ \\
\hline \hline
$k_{1(a)}$ & $0$ &  $2 $ & $2 $ & $4 $ \\
\hline
$k_{2(a)}$ & $0$ &  $2 $ & $0 $ & $2 $ \\
\hline
\end{tabular}
\end{center} 

For $n=3$, the six algebraic eigenvalues are the solutions of the equation
\begin{equation}
E^6 -(248+3c^2)E^4+(4800+240c^2+3c^4)E^2-(23040-1344 c^2-8c^4+c^6)=0 \ .
\end{equation}
As for the case $n=2$ it is apparent that the spectrum of 
the equation is invariant under the reflection
$E\rightarrow -E$.
The three values of $|E|$ are plotted as functions of $c$ in Fig. 1. 
This demonstrates
in particular that a level degeneracy occurs at $c\approx 5$.

The reflection symmetry of the energy eigenvalues $E\rightarrow -E$
can be demonstrated for arbitrary values of $n$ by using
a similar argument as pointed out in \cite{st} for scalar equations.
In the case of (\ref{ham}) the relevant symmetry of the eigenvalue
equation is:
\begin{equation}
y \rightarrow iy \  \ \ , \ \ \phi(E) \rightarrow \sigma_3 \phi(-E) \ .
\end{equation}
Of course, as in the case of \cite{st}, the inclusion of the
$y^4$-term in the potential would spoil this reflection symmetry.

To our knowledge, no QES matrix Schr\"odinger operator is known which preserves 
${\cal P}(n)\oplus {\cal P}(n-\Delta)$ with $\Delta \geq 3$.

\section{Concluding remarks and Outlook}
In this paper we have given a new construction of the realisations
of the super Lie algebra q(2) by means of QES operators. This demonstrates that
the set of operators preserving the vector space
${\cal P}(n-2) \oplus {\cal P}(n)$ is just isomorphic to the 
enveloping algebra of q(2). The quadratic algebra called
${\cal A}(2)$ in \cite{bgk} can be replaced by q(2).

The natural extension of these results would be to study 
$\Delta > 2$. Unfortunately, for generic values of $\Delta$
the combination (\ref{combi}) is limited to $F_{\alpha} = \overline
Q_{\alpha} + c Q_{\alpha}$, ($\alpha = 1,\dots,\Delta + 1$) and no
simplification occurs. For even values of $\Delta$, multiplets having
the same tensorial structure as the $\overline Q$'s can be constructed out
of the $T$'s, e.g. for $\Delta = 4$ we find~:
\begin{equation}
S_{\alpha} = \left( (T^+)^2 \quad , \quad 
             \frac{1}{2} \{ T^+ , T^0 \} \quad , \quad
\frac{1}{3}(2 (T^0)^2 + \frac{1}{2} \{ T^+ , T^- \}) \quad , \quad
\frac{1}{2} \{ T^0 , T^- \} \quad , \quad
(T^-)^2 \right)
\end{equation}
which behaves as a 5-plet (similarly to the $\overline Q_{\alpha}$'s
under the $T$'s). Considering combinations analog to (\ref{combi})
\begin{equation}
F_{\alpha} = \overline Q_{\alpha} + c P_{\alpha} + D S_{\alpha} 
\quad , \quad \alpha = 1, \dots , 5
\end{equation}
we have shown that no values of $c$ and $D$ can be constructed such
that the anti-commutators $\{ F_{\alpha} , F_{\beta} \}$
are polynomials in the bosonic operators.

Another possible extension of these considerations would be the
construction of  finite difference operators preserving 
${\cal P}(n-2) \oplus {\cal P}(n)$. It could be checked then, whether
such operators
obey some deformation of q(2) in a similar way as in \cite{bgk}
where the relevant stucture is the so called 
quommutator deformation
of the Lie superalgebra osp(2,2).\\
\\
\\

{\bf Acknowledgements}
Y. B. gratefully acknowledges the Belgian F.N.R.S. for financial support. 
B.H. was supported by an EPSRC grant.

\newpage
\begin{figure}
\centering
\epsfysize=12cm
\mbox{\epsffile{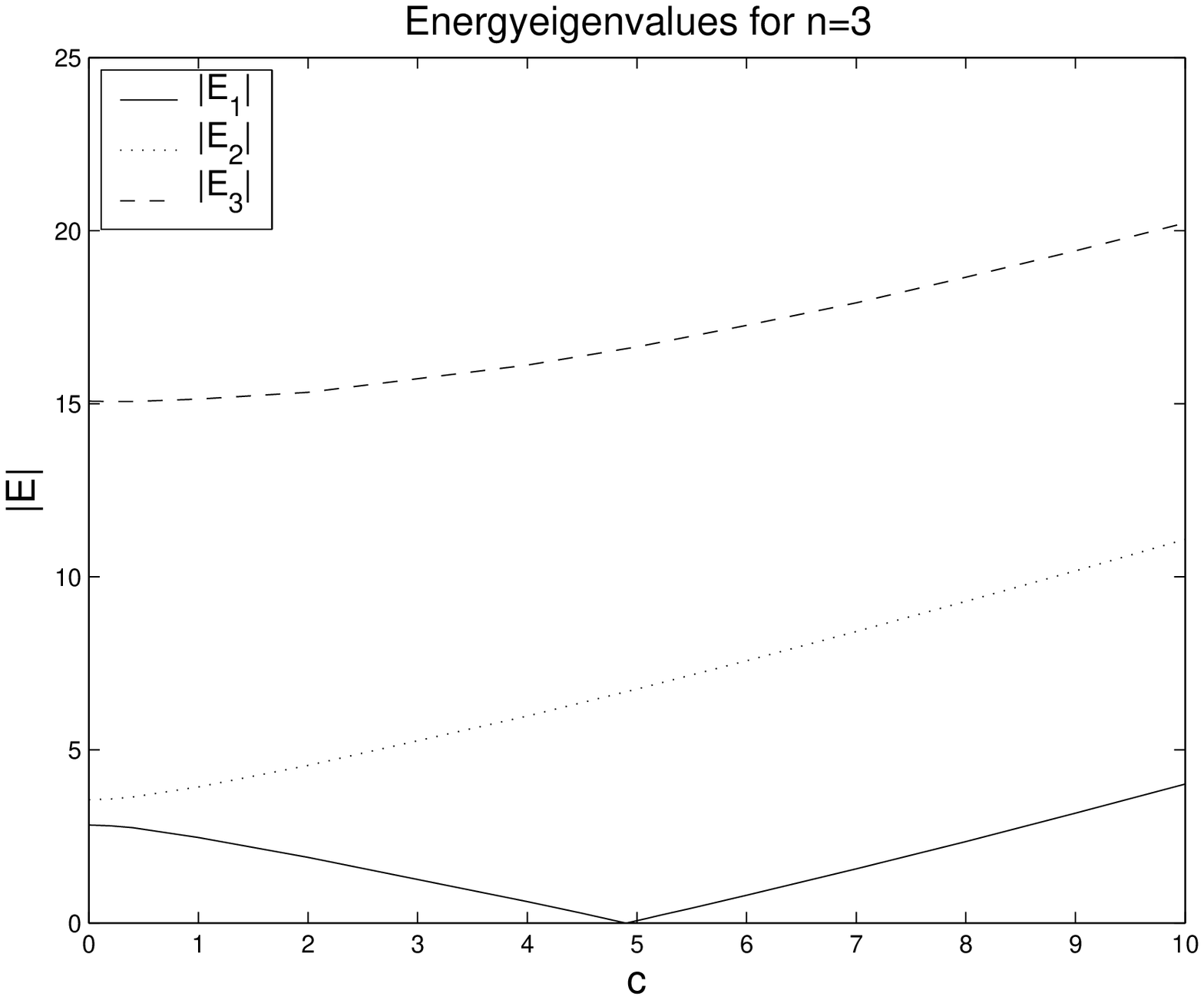}}
\caption{The modulus of the energyeigenvalues is shown for the QES Hamiltonian
presented in Section IV with $n=3$.
 }
\end{figure}
\end{document}